\documentclass[aps,prd,onecolumn,superscriptaddress,nofootinbib,showpacs]{revtex4}
\usepackage{graphicx}
\usepackage{amssymb,amsmath,amsbsy}

\def\iso{\mathchoice{\cong}{\cong}{\isoS}{\cong}}
\def\isoS{\vbox{\baselineskip 0pt  \lineskip 0.5pt
    \ialign{$ \mathsurround=0pt  \scriptstyle \hfil ## \hfil $\crcr
        \sim \crcr = \crcr}}}
\newenvironment{Eqnarray}%
     {\arraycolsep 0.14em\begin{eqnarray}}{\end{eqnarray}}
\def\beqa{\begin{Eqnarray}}
\def\eeqa{\end{Eqnarray}}
\def\beq{\begin{equation}}
\def\eeq{\end{equation}}
\def\eq#1{Eq.~(\ref{#1})}
\def\eqst#1#2{Eqs.~(\ref{#1})--(\ref{#2})}
\def\eqs#1#2{Eqs.~(\ref{#1}) and (\ref{#2})}
\def\Eq#1{Eq.~(\ref{#1})}
\def\eqs#1#2{Eqs.~(\ref{#1}) and (\ref{#2})}

\def\ifmath#1{\relax\ifmmode #1\else $#1$\fi}
\def\half{\tfrac{1}{2}}

\begin{document}
\begin{flushright}
SCIPP-10/01\\[-1mm]
\end{flushright}
\vspace*{1cm}

\title{Basis invariant conditions for supersymmetry in the
  two-Higgs-doublet model}

\author{P.\ M.\ Ferreira}
\affiliation{Instituto Superior de Engenharia de Lisboa,
    Rua Conselheiro Em\'{\i}dio Navarro,
    1900 Lisboa, Portugal}
\affiliation{Centro de F\'{\i}sica Te\'{o}rica e Computacional,
    Faculdade de Ci\^{e}ncias,
    Universidade de Lisboa,
    Av.\ Prof.\ Gama Pinto 2,
    1649-003 Lisboa, Portugal}
\author{Howard E.\ Haber}
\affiliation{Santa Cruz Institute for Particle Physics,
    University of California,
    Santa Cruz, California 95064, USA}
\author{Jo\~{a}o P.\ Silva}
\affiliation{Instituto Superior de Engenharia de Lisboa,
    Rua Conselheiro Em\'{\i}dio Navarro,
    1900 Lisboa, Portugal}
\affiliation{Centro de F\'{\i}sica Te\'{o}rica de Part\'{\i}culas,
    Instituto Superior T\'{e}cnico,
    P-1049-001 Lisboa, Portugal}

\date{\today}

\begin{abstract}
The minimal supersymmetric standard model
involves a rather restrictive Higgs potential
with two Higgs fields.
Recently,
the full set of classes of symmetries allowed
in the most general two Higgs doublet model was identified;
these classes do not include the supersymmetric limit
as a particular class.
Thus,
a physically meaningful definition of the supersymmetric limit
must involve the interaction of the Higgs sector with
other sectors of the theory.
Here we show how one can construct basis
invariant probes of supersymmetry involving both
the Higgs sector and the gaugino-higgsino-Higgs interactions.
\end{abstract}


\maketitle

\section{\label{sec:intro}Introduction}

The Standard Model (SM) of electroweak interactions
has provided an extraordinarily successful description of
currently observed particle physics phenomena.
Nevertheless, there are strong reasons to expect that new physics
beyond the Standard Model must emerge,
ranging from the hierarchy problem and the unification
of all coupling constants, to baryogenesis and dark matter.
One of the leading candidates for physics beyond the SM incorporates
supersymmetry near the scale of electroweak symmetry breaking in order
to provide a natural explanation for the existence of the Higgs boson.
Much attention has been devoted to the
minimal supersymmetric extension of the standard model (MSSM),
which requires two complex Higgs doublets and superpartners for all
Standard Model particles~\cite{MSSM}.

In general, there
is no fundamental reason why the SM should possess only
one complex Higgs doublet. The most well-studied extended Higgs sector
is that of the two-Higgs-doublet model (THDM)~\cite{THDM}.
The scalar potential of the most general THDM involves 14 parameters.
Of these parameters, only eleven combinations are physical, as
three degrees of freedom can be absorbed into a redefinition of the
Higgs fields~\cite{Lav,DavHab}.
This number may be further reduced by imposing some symmetry requirements
on the Higgs Lagrangian.  But,
identifying such symmetries is complicated by the fact that
one may perform a basis transformation of the Higgs fields.
A symmetry that looks simple in one basis may be completely obscured
in another basis.
Hence, it is important to develop basis-invariant signals of such
symmetries, which can identify the
physically meaningful and experimentally accessible
parameters in the theory.
The need to seek basis invariant
observables in models with many Higgs bosons was pointed out by Lavoura and
Silva \cite{LS}, and by Botella and Silva \cite{BS}, stressing
applications to CP violation.
Refs.~\cite{BS,BLS} indicate how to
construct basis invariant quantities in a systematic fashion for any
model, including multi-Higgs-doublet models.
A number of recent articles concerning symmetries and/or basis invariance
in the THDM include Refs.
\cite{Gun,BRSM,GK,DavHab,Haber,oneil,Nishi,NachMani,Ivanov1,Ivanov2,Gerard,FS2,FHS}.

It is remarkable that there are exactly six classes of symmetries
that one may impose on the scalar sector of the most general THDM.
This was shown by Ivanov \cite{Ivanov1} and expanded upon by us
in Ref.~\cite{FHS}.
Since the Higgs sector of the MSSM is a particular case of the THDM,
one would expect that the constraints satisfied by the Higgs sector
of the MSSM would correspond to one of the six classes of symmetries
identified in the scalar sector of the THDM.  
This is \textit{not} the case.
The correct conclusion is that
a physically meaningful definition of the supersymmetric limit
must involve the interaction of the Higgs sector with
other sectors of the supersymmetric theory.
In this article we construct basis-invariant
probes of supersymmetry involving both
the Higgs sector and the gaugino-higgsino-Higgs interactions.

This article is organized as follows.
In section~\ref{sec:notation} we
introduce our notation.
In section~\ref{sec:MSSM}, we
construct the basis invariant quantities that
identify the supersymmetric limit of the scalar
sector of the THDM.
We draw our conclusions in
section~\ref{sec:conclusions}.

\section{\label{sec:notation}The scalar sector of the THDM}

\subsection{The scalar potential}

Let us consider a $SU(2) \otimes U(1)$ gauge theory with
two hypercharge-one Higgs-doublets, denoted by $\Phi_a$,
where $a=1,2$.
The scalar potential may be written as
\begin{eqnarray}
V_H
&=&
m_{11}^2 \Phi_1^\dagger \Phi_1 + m_{22}^2 \Phi_2^\dagger \Phi_2
- \left[ m_{12}^2 \Phi_1^\dagger \Phi_2 + \textrm{h.c.} \right]
\nonumber\\[6pt]
&&
+ \tfrac{1}{2} \lambda_1 (\Phi_1^\dagger\Phi_1)^2
+ \tfrac{1}{2} \lambda_2 (\Phi_2^\dagger\Phi_2)^2
+ \lambda_3 (\Phi_1^\dagger\Phi_1) (\Phi_2^\dagger\Phi_2)
+ \lambda_4 (\Phi_1^\dagger\Phi_2) (\Phi_2^\dagger\Phi_1)
\nonumber\\[6pt]
&&
+ \left[
\tfrac{1}{2} \lambda_5 (\Phi_1^\dagger\Phi_2)^2
+ \lambda_6 (\Phi_1^\dagger\Phi_1) (\Phi_1^\dagger\Phi_2)
+ \lambda_7 (\Phi_2^\dagger\Phi_2) (\Phi_1^\dagger\Phi_2)
+ \textrm{h.c.}
\right],
\label{VH1}
\end{eqnarray}
where $m_{11}^2$, $m_{22}^2$, and $\lambda_1,\lambda_2,\lambda_3,\lambda_4$
are real parameters, and
$m_{12}^2$, $\lambda_5$, $\lambda_6$ and $\lambda_7$
are potentially complex.

An alternative notation,
useful for the construction of invariants
and championed by Botella and Silva \cite{BS} is
\begin{eqnarray}
V_H
&=&
Y_{ab} (\Phi_a^\dagger \Phi_b) +
\tfrac{1}{2}
Z_{ab,cd} (\Phi_a^\dagger \Phi_b) (\Phi_c^\dagger \Phi_d),
\label{VH2}
\end{eqnarray}
where Hermiticity implies
\begin{eqnarray}
Y_{ab} &=& Y_{ba}^\ast,
\nonumber\\
Z_{ab,cd} \equiv Z_{cd,ab} &=& Z_{ba,dc}^\ast.
\label{hermiticity_coefficients}
\end{eqnarray}
One should be very careful when comparing Eqs.~(\ref{VH1})
and (\ref{VH2}) among different authors,
since the same symbol may be used for quantities
that differ by signs, factors of two, or complex conjugation.
Here we follow the definitions of Davidson and Haber
\cite{DavHab}.
With these definitions:
\begin{eqnarray}
Y_{11}=m_{11}^2, &&
Y_{12}=-m_{12}^2,
\nonumber \\
Y_{21}=-(m_{12}^2)^\ast && Y_{22}=m_{22}^2,
\label{ynum}
\end{eqnarray}
and
\begin{eqnarray}
Z_{11,11}=\lambda_1, && Z_{22,22}=\lambda_2,
\nonumber\\
Z_{11,22}=Z_{22,11}=\lambda_3, && Z_{12,21}=Z_{21,12}=\lambda_4,
\nonumber \\
Z_{12,12}=\lambda_5, && Z_{21,21}=\lambda_5^\ast,
\nonumber\\
Z_{11,12}=Z_{12,11}=\lambda_6, && Z_{11,21}=Z_{21,11}=\lambda_6^\ast,
\nonumber \\
Z_{22,12}=Z_{12,22}=\lambda_7, && Z_{22,21}=Z_{21,22}=\lambda_7^\ast.
\label{znum}
\end{eqnarray}

\subsection{Basis transformations}

The scalar potential can be rewritten in terms of new fields $\Phi^\prime_a$,
obtained from the original ones by a simple
(global) basis transformation
\begin{equation}
\Phi_a \rightarrow \Phi_a^\prime = U_{ab} \Phi_b,
\label{basis-transf}
\end{equation}
where $U\in U(2)$ is a $2 \times 2$ unitary matrix.
Under this unitary basis transformation,
the gauge-kinetic terms are unchanged,
but the coefficients $Y_{ab}$ and $Z_{ab,cd}$ are transformed as
\begin{eqnarray}
Y_{ab} & \rightarrow &
Y^\prime_{ab} =
U_{a \alpha}\, Y_{\alpha \beta}\, U_{b \beta}^\ast ,
\label{Y-transf}
\\
Z_{ab,cd} & \rightarrow &
Z^\prime_{ab,cd} =
U_{a\alpha}\, U_{c \gamma}\,
Z_{\alpha \beta,\gamma \delta}\, U_{b \beta}^\ast \, U_{d \delta}^\ast .
\label{Z-transf}
\end{eqnarray}
Thus,
the basis transformations $U$ may be utilized in order to absorb
some of the degrees of freedom of $Y$ and/or $Z$,
which implies that not all parameters of Eq.~(\ref{VH2})
have physical significance.

\subsection{\label{subsec:symmetries}The six classes of symmetries in the THDM}

Symmetries leaving the scalar Lagrangian unchanged
may be of two types.
On the one hand,
one may relate $\Phi_a$ with some unitary transformation
of $\Phi_b$:
\begin{equation}
\Phi_a \rightarrow \Phi_a^S = S_{ab} \Phi_b,
\label{S-transf-symmetry}
\end{equation}
where $S$ is a unitary matrix.
These are known as Higgs Family symmetries, or HF symmetries.
As a result of this symmetry,
\begin{eqnarray}
Y_{a b} & = &
S_{a \alpha}\, Y_{\alpha \beta}\, S_{b \beta}^\ast ,
\label{Y-S}
\\
Z_{ab,cd} & = &
S_{a \alpha}\, S_{c \gamma}\,
Z_{\alpha \beta, \gamma \delta}\, S_{b \beta}^\ast \, S_{d \delta}^\ast .
\label{Z-S}
\end{eqnarray}

On the other hand,
one may relate $\Phi_a$ with some unitary transformation
of $\Phi_b^\ast$:
\begin{equation}
\Phi_a \rightarrow \Phi^{\textrm{GCP}}_a
= X_{a \alpha} \Phi_\alpha^\ast,
\label{GCP}
\end{equation}
where $X$ is an arbitrary unitary matrix.\footnote{The space
coordinates of the fields,
which we have suppressed,
are inverted by a generalized CP transformation.}
These are known as generalized CP symmetries,
or GCP symmetries~\cite{GCP1,GCP2}.
The potential is invariant under this symmetry if and only if
\begin{eqnarray}
Y_{ab}^\ast
&=&
X_{\alpha a}^\ast Y_{\alpha \beta} X_{\beta b}
\nonumber\\
Z_{ab,cd}^\ast
&=&
X_{\alpha a}^\ast X_{\gamma c}^\ast
Z_{\alpha \beta, \gamma \delta} X_{\beta b} X_{\delta d}.
\label{YZ-CPtransf}
\end{eqnarray}

Under the basis transformation of Eq.~(\ref{basis-transf}),
the specific forms of the HF and GCP symmetries are altered,
respectively, as follows:
\begin{eqnarray}
S^\prime &=& U S U^\dagger ,
\label{S-prime}
\\
X^\prime &=& U X U^\top.
\label{X-prime}
\end{eqnarray}
Hence, a basis-invariant treatment is critical for
distinguishing between two potentially different symmetries.

Of course,
one may combine several HF symmetries and/or GCP symmetries.
Ivanov \cite{Ivanov1} has proved that,
whatever combination one chooses,
one will end up in one of six distinct classes of symmetries.
In a recent article we have clarified this issue showing
how to construct such classes
with simple examples \cite{FHS}.
The result is shown in Table~\ref{master1}.
%
\begin{table}[ht!]
\caption{Impact of the symmetries on the coefficients
of the Higgs potential in a specified basis.
See Ref.~\cite{FHS} for more details.}
\begin{ruledtabular}
\begin{tabular}{ccccccccccc}
symmetry & $m_{11}^2$ & $m_{22}^2$ & $m_{12}^2$ &
$\lambda_1$ & $\lambda_2$ & $\lambda_3$ & $\lambda_4$ &
$\lambda_5$ & $\lambda_6$ & $\lambda_7$ \\
\hline
$Z_2$ &   &   & 0 &
   &  &  &  &
   & 0 & 0 \\
$U(1)$ &  &  & 0 &
 &  & &  &
0 & 0 & 0 \\
$SO(3)$ &  & $ m_{11}^2$ & 0 &
   & $\lambda_1$ &  & $\lambda_1 - \lambda_3$ &
0 & 0 & 0 \\
\hline
CP1 &  &  & real &
 & &  &  &
real & real & real \\
CP2 &  & $m_{11}^2$ & 0 &
  & $\lambda_1$ &  &  &
   &  & $- \lambda_6$ \\
CP3 &  & $m_{11}^2$ & 0 &
   & $\lambda_1$ &  &  &
$\lambda_1 - \lambda_3 - \lambda_4$ (real) & 0 & 0 \\
\end{tabular}
\end{ruledtabular}
\label{master1}
\end{table}
%

Five of the symmetry classes may be imposed by the following single requirements:
\begin{eqnarray}
Z_2: &\hspace{1cm} &
S =
\left(
\begin{array}{cc}
1 & 0\\
0 & -1
\end{array}
\right),
\\
U(1): &\hspace{1cm} &
S =
\left(
\begin{array}{cc}
e^{i \alpha} & 0\\
0 & e^{-i \alpha}
\end{array}
\right)_{\alpha\neq \pi/2},
\\
CP1: &\hspace{1cm} &
X =
\left(
\begin{array}{cc}
1 & 0\\
0 & 1
\end{array}
\right),
\\
CP2: &\hspace{1cm} &
X =
\left(
\begin{array}{cc}
0 & 1\\
-1 & 0
\end{array}
\right),
\\
CP3: &\hspace{1cm} &
X =
\left(
\begin{array}{cc}
\cos{\theta} & \sin{\theta}\\
-\sin{\theta} & \cos{\theta}
\end{array}
\right).
\end{eqnarray}
Here $0<\alpha<\pi$ (but $\alpha\neq \pi/2$, since the case of $\alpha=\pi/2$
corresponds to the $Z_2$ symmetry), and $0<\theta<\pi/2$.
Invariance under the full $U(2)$
global symmetry\footnote{The $SO(3)$
Higgs flavor symmetry listed in Table~\ref{master1} is orthogonal
to the $U(1)_{\rm Y}$ hypercharge invariance (under which the THDM
potential is always invariant).  In the $SO(3)$-symmetric case, 
the \textit{full} Higgs flavor
symmetry group is $U(2)\iso SO(3)\otimes 
U(1)_{\rm Y}$~\cite{FHS}. \label{fn}}
is obtained by requiring the invariance of the
scalar potential under Eq.~(\ref{S-transf-symmetry}),
for \textit{all} unitary matrices $S$.

\section{\label{sec:MSSM}Basis invariant probes of the MSSM}

\subsection{\label{sec:Higgs_MSSM}The Higgs sector of the MSSM}

The Higgs potential of the MSSM (prior to including
soft-supersymmetry-breaking dimension-two squared-mass terms)
is a particular case of Eq.~(\ref{VH1}), with
\begin{eqnarray}
m_{11}^2 &=& m_{22}^2,
\label{MSSM_m11_m22}
\\
m_{12}^2 &=& 0,
\label{MSSM_m12}
\\
\lambda_1 = \lambda_2 &=&
\tfrac{1}{4} (g^2 + {g^\prime}^2),
\label{MSSM_L1_L2}
\\
\lambda_3 &=&
\tfrac{1}{4} (g^2 - {g^\prime}^2),
\label{MSSM_L3}
\\
\lambda_4 &=&
- \tfrac{1}{2} g^2,
\label{MSSM_L4}
\\
\lambda_5=\lambda_6=\lambda_7 &=& 0,
\label{MSSM_L5_L6_L7}
\end{eqnarray}
where $g$ and $g^\prime$ are the $SU(2)_L$ and $U(1)_Y$
gauge coupling constants,
respectively.
In this case,
Eqs.~(\ref{ynum}) and (\ref{znum}) become
\begin{eqnarray}
Y_{11}=Y_{22}, &&
Y_{12}=Y_{21}=0,
\label{yESL}
\end{eqnarray}
and
\begin{eqnarray}
Z_{11,11}=\lambda_1, && Z_{22,22}=\lambda_1,
\nonumber\\
Z_{11,22}=Z_{22,11}=\lambda_3, && Z_{12,21}=Z_{21,12}=- \lambda_1 - \lambda_3,
\label{zESL}
\end{eqnarray}
with $\lambda_1$ given by Eq.~(\ref{MSSM_L1_L2}),
$\lambda_3$ given by Eq.~(\ref{MSSM_L3}),
and all other components of the $Z$ tensor equal to zero.

Comparing Eqs.~(\ref{MSSM_m11_m22})--(\ref{MSSM_L5_L6_L7})
with Table~\ref{master1},
we see that these requirements are almost the same as in
the THDM with the full $U(2)$ flavor symmetry\footnotemark[\arabic{footnote}].
The difference is that the $U(2)$-symmetric case implies
$\lambda_4 = \lambda_1 - \lambda_3$,
while the supersymmetry limit
implies $\lambda_4 = - \lambda_1 - \lambda_3$.
As shown by Ivanov \cite{Ivanov1} and by us \cite{FHS},
the former relation can come from a symmetry requirement
that exclusively involves
the Higgs potential, while the latter relation
cannot. In particular, there are no
basis changes one can perform on the THDM to obtain, from one of the six
symmetries listed in table~\ref{master1}, the SUSY condition
$\lambda_4 = - \lambda_1 - \lambda_3$.

This can also be seen by examining the renormalization
group equations that control the evolution of the
$\lambda_i$.
Here, we focus only on those terms arising from the Higgs
potential and the gauge couplings.\footnote{In order to include
fermions in the analysis, one would have to investigate the
constraints of the THDM symmetries on the Higgs-fermion Yukawa
couplings.}
The relevant expressions can be found,
for example,
in Refs.~\cite{CEL,HH,GL,FJ}.
Using $\lambda_1=\lambda_2$ and $\lambda_5=\lambda_6=\lambda_7=0$,
we find
\begin{eqnarray}
{\cal D} \lambda_1 &=&
6 \lambda_1^2 + 2 \lambda_3^2 + \lambda_4^2 + 2 \lambda_3 \lambda_4
- \tfrac{1}{2} \left( 9 g^2 + 3 {g^\prime}^2 \right) \lambda_1
+ \tfrac{1}{8} \left( 9 g^4 + 6 g^2 {g^\prime}^2 + 3{g^\prime}^4 \right),
\nonumber\\
{\cal D} \lambda_3 &=&
2 \lambda_3^2 + \lambda_4^2 + 2 \lambda_1 (3 \lambda_3 + \lambda_4)
- \tfrac{1}{2} \left( 9 g^2 + 3 {g^\prime}^2 \right) \lambda_3
+ \tfrac{1}{8} \left( 9 g^4 - 6 g^2 {g^\prime}^2 + 3{g^\prime}^4 \right),
\nonumber\\
{\cal D} \lambda_4 &=&
2 \lambda_4^2 + 2 \lambda_1 \lambda_4 + 4 \lambda_3 \lambda_4
- \tfrac{1}{2} \left( 9 g^2 + 3 {g^\prime}^2 \right) \lambda_4
+ \tfrac{3}{2} g^2 {g^\prime}^2, \label{D134}
\end{eqnarray}
where ${\cal D} = 16 \pi^2 \mu (d/d\mu)$,
and ${\cal D} \lambda_5 = {\cal D} \lambda_6 = {\cal D} \lambda_7 = 0$.
Hence,
given the constraints
$\lambda_1=\lambda_2$ and $\lambda_5=\lambda_6=\lambda_7=0$,
\begin{eqnarray}
{\cal D} (\lambda_4 + \lambda_3 - \lambda_1 ) &=&
\tfrac{1}{2}
(\lambda_4 + \lambda_3 - \lambda_1 )(12 \lambda_1 + 4 \lambda_4 - 9 g^2 - 3 {g^\prime}^2)
\label{D_O3}
\\*[2mm]
{\cal D} (\lambda_4 + \lambda_3 + \lambda_1 ) &=&
2 \left(3 \lambda _1^2+\left(3 \lambda _3+2 \lambda _4\right) \lambda _1+2 \lambda _3^2+2
   \lambda _4^2+3 \lambda _3 \lambda _4\right)
\nonumber\\
&&
- \tfrac{1}{2} \left( 9 g^2 + 3 {g^\prime}^2 \right)
\left(\lambda_4 + \lambda_3 + \lambda_1 \right)
+ \tfrac{1}{4} \left( 9 g^4 + 6 g^2 {g^\prime}^2 + 3{g^\prime}^4 \right).
\label{D_MSSM}
\end{eqnarray}
The first equation vanishes if $\lambda_4 = \lambda_1 - \lambda_3$;
the second does not vanish if $\lambda_4 = - \lambda_1 - \lambda_3$.
That is, the condition
$\lambda_4 = \lambda_1 - \lambda_3$ is renormalization group invariant,
whereas the condition $\lambda_4 = - \lambda_1 - \lambda_3$ is not.
Note that we have not yet imposed the specific relations between the $\lambda_i$ and
the gauge couplings required by the MSSM.  If we impose the MSSM
constraints specified by Eqs.~(\ref{MSSM_L1_L2})--(\ref{MSSM_L4})
on the right hand side of \eq{D_MSSM}, we obtain
\beq \label{RGINV}
{\cal D} (\lambda_4 + \lambda_3 + \lambda_1 ) =
3 g^4 + 2 g^2 {g^\prime}^2 + {g^\prime}^4\,,
\eeq
i.e., $\lambda_1 = - \lambda_3 - \lambda_4$ is still not RGE invariant.

The latter result is not unexpected.  After all, the gauge
boson--Higgs boson sector considered by itself can never be
supersymmetric, as the corresponding superpartners are not included.
Consequently, the SUSY limit of the gauge
boson--Higgs boson sector
can only be defined in a manner invariant under
Higgs basis changes if the corresponding gaugino and higgsino
superpartners are taken into account.  The gaugino and higgsino interactions
generate additional terms on the right hand side of \eq{D134}.
In the supersymmetric limit, these effects yield~\cite{HH}
\beqa
\delta_{\rm SUSY}(\mathcal{D}\lambda_1)&=&-\tfrac{5}{2}g^4-g^2 g^{\prime\,2}
-\tfrac{1}{2}g^{\prime\,4}\,,\nonumber \\
\delta_{\rm SUSY}(\mathcal{D}\lambda_3)&=&-\tfrac{5}{2}g^4+g^2 g^{\prime\,2}
-\tfrac{1}{2}g^{\prime\,4}\,,\nonumber \\
\delta_{\rm SUSY}(\mathcal{D}\lambda_4)&=& 2g^4-2g^2 g^{\prime\,2}\,.
\label{deltaD}
\eeqa
Hence,
$$
\delta_{\rm SUSY}\{\mathcal{D}(\lambda_4+\lambda_3+\lambda_1)\}=
-(3 g^4 + 2 g^2 {g^\prime}^2 + {g^\prime}^4)\,.
$$
Indeed, when the latter is added to \eq{RGINV}, we see that
${\cal D} (\lambda_4 + \lambda_3 + \lambda_1 )=0$ as expected.
Thus, the SUSY relation $\lambda_1=-\lambda_3-\lambda_4
=\tfrac{1}{4}(g^2+g^{\prime\,2})$
is renormalization group invariant when all the
Higgs/higgsino/gauge/gaugino interactions are included.

\subsection{\label{sec:necessary}Basis-independent conditions for
the MSSM Higgs potential that are necessary but not sufficient}

Based on the arguments of section~\ref{sec:Higgs_MSSM}, no basis-invariant
conditions constructed solely from the $Y_{ab}$ and $Z_{ab,cd}$ exist
that can guarantee that \eqst{MSSM_m11_m22}{MSSM_L5_L6_L7} 
are satisfied for some choice of basis.  Nevertheless, we can
establish a weaker invariant condition that is necessary (although not
sufficient) for the existence of a supersymmetric THDM scalar
potential.\footnote{We thank the anonymous referee for encouraging
us to consider the necessary symmetry constraints that govern the MSSM Higgs
scalar potential.}

Consider a basis choice in which:
\beq \label{weaker}
m_{11}^2=m_{22}^2\,,\qquad \lambda_1=\lambda_2\,, \quad {\rm and}
\quad m_{12}^2=\lambda_5=\lambda_6=\lambda_7=0\,.
\eeq
Clearly \eq{weaker} is satisfied by the MSSM Higgs potential in the
standard MSSM basis choice for the Higgs fields.  However, \eq{weaker}
is not sufficient since the condition $\lambda_4=-\lambda-\lambda_3$
is imposed, nor are the conditions relating the quartic Higgs
couplings to gauge couplings imposed.  Nevertheless, suppose that one could
establish a basis-independent condition that was equivalent to 
the statement that a basis exists in which \eq{weaker} is satisfied.
Such a basis-independent condition would then be a necessary condition
for a supersymmetric Higgs sector.\footnote{The additional conditions
necessary to establish necessary \textit{and} sufficient
basis-independent probes of supersymmetry require consideration of the
gaugino-higgsino-Higgs interactions.  This is the subject of 
section~\ref{sec:gaugino_Higgs_MSSM} that follows.}

The conditions of \eq{weaker} do not match any single
condition listed in Table~\ref{master1}.  Nevertheless, given
a scalar potential whose parameters satisfy \eq{weaker}, one can
\textit{always} transform to a new basis that explicitly 
satisfies the CP3-symmetry conditions of Table~\ref{master1},
and vice versa.  That is,
a basis-invariant characterization of a CP3-symmetric THDM would
guarantee that some basis exists in which \eq{weaker} holds.
It follows that 
\textit{the Higgs sector of the MSSM is a CP3-symmetric THDM.}

The proof of the assertions above have been given in Ref.~\cite{FHS}.
For completeness, we review the arguments here.  We first define
the discrete flavor symmetry $\Pi_2$ which corresponds to a THDM scalar
potential that is symmetric under the interchange of the two Higgs
fields.  In this case, the scalar potential parameters satisfy:
\beq \label{pi2}
m_{11}^2=m_{22}^2\,,\qquad m_{12}^2~\text{real}\,,\qquad
\lambda_1=\lambda_2\,,\qquad \lambda_5~\text{real}\,,\quad {\rm and}
\quad \lambda_6=\lambda_7\,.
\eeq
In Ref.~\cite{FHS}, we showed that given a $\Pi_2$-symmetric
scalar potential, there exists another basis which is $Z_2$-symmetric.
Consequently, we did not list $\Pi_2$ as a separate symmetry in
Table~\ref{master1}.  However, if we simultaneously impose $\Pi_2$
and $U(1)$ \textit{in the same basis}, then one easily sees that
the conditions of \eq{weaker} are satisfied.  The same conclusion also
follows if we simultaneously impose $U(1)$ and CP2 in the
same basis.  Moreover, starting from a scalar potential in 
which the conditions of \eq{weaker} are satisfied, a
basis transformation can be found 
(see Ref.~\cite{FHS} for the details) such that 
$\lambda_5=\lambda_1-\lambda_3-\lambda_4$ in the new basis, as required in 
the CP3-symmetric THDM [cf.~Table~\ref{master1}].  

Finally, we indicate the basis-independent conditions that guarantee
that a basis exists in which the CP3-symmetric conditions are
satisfied. Defining $Z_{ab}^{(1)}\equiv Z_{ac,cb}$, we first require
that~\cite{DavHab}
\beq \label{erps}
 Y_{ab}=m_{11}^2\delta_{ab}\quad \text{and} \quad
{\rm Tr}[Z^{(1)}]^2=\half({\rm Tr}~Z^{(1)})^2\,.
\eeq
If \eq{erps} is satisfied, then
\beq \label{erps2}
m_{11}^2=m_{22}^2\,,\qquad m_{12}^2=0\,,\qquad
\lambda_1=\lambda_2\,,\qquad \lambda_7=-\lambda_6\,,
\eeq
must hold in \textit{all basis choices}.  This is the so-called
exceptional region of parameter space (ERPS) of the THDM.
In Ref.~\cite{FHS}, we then 
constructed a second invariant quantity $D$ built out of the $Y_{ab}$
and $Z_{ab,cd}$
with the following property: if $D=0$ in the ERPS,
then there exists a basis for the scalar fields such that the
CP3-conditions of Table~\ref{master1} are satisfied.\footnote{The 
explicit expression for $D$ is rather
complicated and can be found in eqs.~(39)--(41) and (44) of Ref.~\cite{FHS}.} 
Thus, the conditions for the ERPS plus $D=0$ provide necessary
(although not sufficient) invariant conditions that must be satisfied
by the MSSM Higgs scalar potential.

\subsection{\label{sec:gaugino_Higgs_MSSM}The gaugino-higgsino-Higgs interactions}

In the MSSM, the tree-level Lagrangian describing the interactions of
the gauginos with
the Higgs-doublets may be written as
\beq \label{susygaugino}
{\cal L}_{\textrm{gaugino-Higgs}}^{\textrm{MSSM}}=\mu\epsilon_{ij}\psi^i_{H_D}
\psi^j_{H_U}+
\frac{ig}{\sqrt{2}}
\lambda^\alpha \tau^\alpha_{ij}\left(\psi^j_{H_U}\Phi_2^{i\,\dagger}
+\epsilon^{ik}\psi^j_{H_D}\Phi_1^k\right)
+\frac{ig'}{\sqrt{2}}\lambda^\prime\left(\psi^i_{H_U}\Phi_2^{i\,\dagger}
-\epsilon^{ik}\psi^i_{H_D}\Phi_1^k\right)+{\rm h.c.}\,,
\eeq
where $\lambda^\alpha$ and $\lambda'$ are the two-component spinor gaugino
fields that are superpartners to the SU(2) and U(1)-hypercharge gauge
bosons, and $\psi_{H_D}$ and $\psi_{H_U}$ are, respectively, the
hypercharge $-1$ and hypercharge $+1$ weak doublet
two-component spinor higgsino fields.  The indices $i$, $j$ and $k$
label the two components of the weak doublet, and the index~$\alpha$ is the
adjoint index of the SU(2) gaugino field.  We have included a
supersymmetric Majorana mass term for the two-component higgsino
fields, which defines the parameter $\mu$.  As usual, $\epsilon^{12}=
-\epsilon^{21}=+1$ and $\epsilon^{11}=\epsilon^{22}=0$.

If we relax the constraints imposed by supersymmetry,
the coupling strengths of the gaugino-higgsino-Higgs interaction above are no
longer constrained to be gauge couplings as in \eq{susygaugino}.
Moreover,
four additional dimension-four interaction terms are possible, consistent with
SU(2)$\times$U(1) gauge invariance.  These terms
are the so-called ``wrong-Higgs'' couplings of Ref.~\cite{HM},
and are obtained from
those of \eq{susygaugino} by interchanging $\Phi_1$ and $\Phi_2$.
In our analysis below, we consider the most general dimension-four
gauge invariant couplings between the gaugino, higgsino and Higgs
fields.  We shall write these couplings in a form that is manifestly
independent of the choice of basis for the Higgs fields:
\beq \label{gensusygaugino}
{\cal L}_{\textrm{gaugino-Higgs}}=\frac{i}{\sqrt{2}}
\lambda^\alpha \tau^\alpha_{ij}\left(\psi^j_{H_U}f_U^a\Phi_a^{i\,\dagger}
+\epsilon^{ik}\psi^j_{H_D}f_D^{a\,*}\Phi_a^k\right)
+\frac{i}{\sqrt{2}}\lambda^\prime\left(\psi^i_{H_U}
f^{\prime\,a}_U\Phi_a^{i\,\dagger}
-\epsilon^{ik}\psi^i_{H_D}f^{\prime\,a\,*}_D\Phi_a^k\right)+{\rm h.c.}\,,
\eeq
where the couplings $f^a_U$, $f^a_D$, $f^{\prime\,a}_U$, and
$f^{\prime\,a}_D$ transform covariantly
under a Higgs basis U(2)-transformation,\footnote{Note
that the global U(1) Higgs
flavor transformation corresponding to $\Phi_a\to e^{i\chi}\Phi_a$ ($a=1,2$)
is distinguished from the global U(1) hypercharge transformation, since
the higgsino fields \textit{do not} transform under the rephasing of
the Higgs fields.}
\beq
f^a_{U,D}\to U_{ab}f^b_{U,D}\,,\qquad\quad
f^{\prime\,a}_{U,D}\to U_{ab}f^{\prime\,b}_{U,D}\,.
\eeq
In the supersymmetric limit, there is a natural choice of basis for
the Higgs fields, henceforth called the \textit{SUSY basis}, in which:
\beq \label{spbasis}
f_U^a=g\delta^{a2}\,,\qquad\quad
f_D^{a\,*}=g\delta^{a1}\,,\qquad\quad
f_U^{\prime\,a}=g'\delta^{a2}\,,\qquad\quad
f_D^{\prime\,a\,*}=g'\delta^{a1}\,.
\eeq
In particular, in the SUSY basis, the so-called ``wrong-Higgs
interactions'' of Ref.~\cite{HM} are absent in the supersymmetric
limit.  However, under a general
Higgs basis transformation, the supersymmetric
gaugino-higgsino-Higgs Lagrangian will transform into a linear
combination of supersymmetric and wrong-Higgs interaction terms.
Thus, in a generic basis choice for the Higgs fields,
the supersymmetry is not manifest.
One of the
goals of this section is to determine a set of
basis-independent conditions that guarantees the existence of a basis
choice in which \eq{spbasis} is satisfied.  Such basis-independent
conditions would constitute an invariant signal for manifestly
supersymmetric Higgs interactions.

The couplings $f^a_U$, $f_D^a$, $f^{\prime\,a}_U$, and
$f^{\prime\,a}_D$ are complex vectors that live in the two-dimensional Higgs
flavor space.  It is convenient to define the corresponding unit
vectors, $\hat f^a\equiv f^a/|f|$,
where $|f|\equiv (f^{a\,\ast} f^a)^{1/2}$ is the length of the complex
vector $f^a$.
Next, we introduce vectors
that are orthogonal to $f^a_U$, $f_D^a$, $f^{\prime\,a}_U$, and
$f^{\prime\,a}_D$, respectively,
\beqa
\hat g_U^a&\equiv&\hat f_U^{b\,*} \epsilon^{ba}\,,\qquad\quad
\hat g_D^a\equiv \hat f_D^{b\,*} \epsilon^{ba}\,,\\
\hat g_U^{\prime\,a}&\equiv&\hat f_U^{\prime\,b\,*} \epsilon^{ba}\,,\qquad\,\,\,
\hat g_D^{\prime\,a}\equiv \hat f_D^{\prime\,b\,*} \epsilon^{ba}\,.
\eeqa
These are pseudo-vectors with respect to U(2) Higgs basis
transformations,\footnote{Starting from the transformation law
$\hat f^a\to U_{ab}\hat f^b$, where $U^\dagger=U^{-1}$, it follows
that $\hat g^c\to U^{-1}_{ab}g^d\epsilon^{ad}\epsilon^{bc}$.  If
we now recognize that $U^{-1}_{ab}\epsilon^{ad}={\rm det}(U^{-1})
U_{cd}\epsilon^{bc}$, the results of \eq{gtransform} easily follow.}
\beq \label{gtransform}
\hat g^a_{U,D}\longrightarrow ({\rm det}~U)^{-1}U_{ab}\,
\hat g^b_{U,D}\,,\qquad\quad
\hat g^{\prime\,a}_{U,D}\longrightarrow ({\rm det}~U)^{-1}U_{ab}\,
\hat g^{\prime\,b}_{U,D}\,,
\eeq
due to the appearance of the complex phase, ${\rm det}~U$.

We now define U(2)-invariant, hypercharge-one Higgs fields
as follows:
\beqa
H_U& \equiv &\hat f_U^{a\,*}\Phi_a\,,\qquad\qquad
H'_U\equiv \hat f_U^{\prime\,a\,*}\Phi_a\,,\label{invH1}\\[6pt]
\widetilde H_D& \equiv & \hat f_D^{a\,*}\Phi_a\,,\qquad\qquad
\widetilde H'_D\equiv \hat f_D^{\prime\,a\,*}\Phi_a\,.\label{invH2}
\eeqa
One can also define a corresponding set of hypercharge $-1$ fields,
e.g.,
\beq \label{invH3}
H_D^i\equiv\epsilon^{ij}\widetilde H_D^{j\,\dagger}\,,\qquad\qquad
H_D^{\prime\,i}\equiv\epsilon^{ij}\widetilde H_D^{\prime\,j\,\dagger}\,.
\eeq
It is also convenient to define U(2) pseudo-invariant Higgs fields
(denoted by calligraphic fonts),
\beqa
\mathcal{H}_U& \equiv &\hat g_U^{a\,*}\Phi_a\,,\qquad\qquad
\mathcal{H}^\prime_U\equiv \hat g_U^{\prime\,a\,*}\Phi_a\,,\label{invH4} \\[6pt]
\widetilde{\mathcal{H}}_D& \equiv & \hat g_D^{a\,*}\Phi_a\,,\qquad\qquad
\widetilde{\mathcal{H}}^\prime_D\equiv \hat g_D^{\prime\,a\,*}\Phi_a\,.\label{invH5}
\eeqa
It then follows that:
\beqa
\Phi_a&=& \hat f_U^a H_U +\hat g_U^a \mathcal{H}_U=
\hat f_D^a \widetilde{H}_D +\hat g_D^a \widetilde{\mathcal{H}}_D
\label{phia1} \\[6pt]
&=& \hat f_U^{\prime\,a} H'_U +\hat g_U^{\prime\,a} \mathcal{H}'_U=
\hat f_D^{\prime\,a} \widetilde{H}'_D
+\hat g_D^{\prime\,a} \widetilde{\mathcal{H}}'_D\,,\label{phia}
\eeqa
after using $\hat f_{U,D}^a \hat f_{U,D}^{a\,*}=
\hat g_{U,D}^a \hat g_{U,D}^{a\,*}=1$ and
$\hat g_{U,D}^a \hat f_{U,D}^{a\,*}=\hat f_{U,D}^a \hat g_{U,D}^{a\,*}=0$.

There is some motivation for this proliferation of Higgs field
definitions.  In particular, as we show later in \eqst{bu}{bdp}, the
choices of
\beq \label{four}
\{\mathcal{H}_U\,,\,H_U\}\,,\,
\{\mathcal{H}^\prime_U\,,\,H^\prime_U\}\,,\,
\{\widetilde{H}_D\,,\,\widetilde{\mathcal{H}}_D\}\,,\,\,\,{\rm  and}\,\,\,
\{\widetilde{H}^\prime_D\,,\,\widetilde{\mathcal{H}}^\prime_D\}\,,
\eeq
correspond to four different
basis choices for the hypercharge-one Higgs doublet fields.

One can express the gaugino-higgsino-Higgs interaction Lagrangian in
a manifestly U(2)-invariant form.  For example, using the definitions
of the invariant Higgs fields $H_U$, $H_D$, $H_U^\prime$ and
$H_D^\prime$ [defined by \eqs{invH1}{invH3}],
\eq{gensusygaugino} can be rewritten as:
\beq \label{gensusygauginoinv}
{\cal L}_{\textrm{gaugino-Higgs}}=\frac{i}{\sqrt{2}}
\lambda^\alpha \tau^\alpha_{ij}\left(|f_U|\psi^j_{H_U}H_U^{i\,\dagger}
+|f_D|\psi^j_{H_D}H_D^{i\dagger}\right)
+\frac{i}{\sqrt{2}}\lambda^\prime\left(|f^\prime_U|\psi^i_{H_U}
H_U^{\prime\,i\,\dagger}
-|f_D^\prime|\psi^i_{H_D}H_D^{\prime\,i\,\dagger}\right)+{\rm h.c.}
\eeq
However, this form is not particularly useful outside of the
supersymmetric limit, since $\{H_U\,,\,\widetilde H_D\}$
and  $\{H^\prime_U\,,\,\widetilde H^\prime_D\}$ are
\textit{not} orthogonal pairs of hypercharge-one Higgs doublet
fields in the general case.
Of course, one can always rewrite \eq{gensusygaugino} in terms of one
of the four basis choices of hyper-charge one doublet Higgs fields
listed in \eq{four} by inserting the appropriate form for $\Phi_a$
given in \eqs{phia1}{phia} into \eq{gensusygaugino}.

\subsection{\label{sec:ESL}Basis-invariant probes of the
supersymmetric Higgs interactions}

Supersymmetry imposes strong constraints on the scalar Higgs potential
and the gaugino-higgsino-Higgs interactions.  These constraints must involve
basis-independent combinations of the scalar potential parameters
$Y_{ab}$, $Z_{abcd}$, and the gaugino-higgsino-Higgs couplings
$f^a_U$, $f_D^a$, $f^{\prime\,a}_U$, $f^{\prime\,a}_D$.  It is
straightforward to find the necessary relations.  First we exhibit
the basis invariant relations that enforce supersymmetric
gaugino-higgsino-Higgs couplings:

\begin{eqnarray}
f_U^a f_D^{a \ast} = 0,
&\hspace{1cm}&
f_U^a f_D^{\prime a \ast} = 0,
\nonumber\\
f_U^{\prime a} f_D^{a \ast} = 0,
&\hspace{1cm}&
f_U^{\prime a} f_D^{\prime a \ast} = 0,
\nonumber\\
f_U^a f_U^{a \ast} = f_D^a f_D^{a \ast}= g^2,
&\hspace{1cm}&
f_U^{\prime a} f_U^{\prime a \ast}
= f_D^{\prime a} f_D^{\prime a \ast} = g^{\prime\,2}\,,
\nonumber \\
f_U^a f_U^{\prime\,a \ast} = gg^\prime,
&\hspace{1cm}&
f_D^{a \ast} f_D^{\prime a} = gg^{\prime}\,.
\label{f_ESL}
\end{eqnarray}
%
%
%

To establish U(2)-invariant
conditions that enforce a supersymmetric scalar Higgs
potential, we first construct basis-independent quantities that
involve both the scalar potential parameters and the gaugino-higgsino-Higgs
couplings.  For example,\footnote{Assuming \eq{f_ESL} is satisfied, it is
not necessary to construct additional invariants that involve
$f^\prime_U$ and $f^\prime_D$.}
\begin{eqnarray}
{\cal Y}_{DD} &=& \hat f_D^a\, \hat f_D^{b \ast}\ Y_{ab},
\qquad\qquad
{\cal Y}_{UU} = \hat f_U^a\, \hat f_U^{b \ast}\ Y_{ab},
\nonumber\\
{\cal Y}_{DU} &=& \hat f_D^a\, \hat f_U^{b \ast}\ Y_{ab},
\qquad\qquad
{\cal Y}_{UD} = \hat f_U^a\, \hat f_D^{b \ast}\ Y_{ab},
\label{calY}
\end{eqnarray}
provide basis-invariant
quantities involving the quadratic coefficients of the Higgs potential.
Likewise,
\begin{equation}
{\cal Z}_{\alpha \beta, \gamma \delta} =
\hat f_\alpha^a\, \hat f_\beta^{b \ast}\,
\hat f_\gamma^c\, \hat f_\delta^{c \ast}\, Z_{ab,cd}\,,
\label{calZ}
\end{equation}
where the indices $\alpha$, $\beta$, $\gamma$, and $\delta$
can take the values $D$ or $U$, provide basis-invariant quartic
coefficients for the Higgs potential.

Evaluating the invariant quantities introduced in \eqs{calY}{calZ}
in the supersymmetric basis defined by
Eqs.~(\ref{yESL}), (\ref{zESL})  and (\ref{spbasis}),
it follows that
\beq
{\cal Y}_{DD} = {\cal Y}_{UU},\qquad\qquad
{\cal Y}_{DU}  =  {\cal Y}_{UD} = 0,
\label{calY_ESL}
\eeq
and
\begin{eqnarray}
{\cal Z}_{DD,DD}&=& {\cal Z}_{UU,UU}  = \tfrac{1}{4}\left[ f_U^{a\,\ast} f_U^a+
 f_D^{a\,\ast} f_D^a\right],
\nonumber\\
{\cal Z}_{DD,UU} &=& {\cal Z}_{UU,DD} = \tfrac{1}{4}
\left[ f_U^{a\,\ast} f_U^a - f_U^{\prime\,a\,\ast} f_U^{\prime\,a}\right],
\nonumber\\
{\cal Z}_{DU,UD} &=& {\cal Z}_{UD,DU} = - {\cal Z}_{DD,DD} - {\cal Z}_{DD,UU},
\nonumber\\*[2mm]
{\cal Z}_{DU,DU} &=& {\cal Z}_{UD,UD} = 0, \nonumber \\
{\cal Z}_{UD,DD}&=&{\cal Z}_{DU,DD}={\cal Z}_{DD,UD}={\cal Z}_{DD,DU}=0,\nonumber\\
{\cal Z}_{DU,UU}&=&{\cal Z}_{UD,UU}={\cal Z}_{UU,DU}={\cal Z}_{UU,UD}=0.
\label{calZ_ESL}
\end{eqnarray}
%
Since these equations are basis invariant,
they must hold in any theory made up of the MSSM fields
with exact supersymmetry,
regardless of the exact basis choice made for the Higgs fields.

That is, independently of the choice of basis for the Higgs fields,
the supersymmetric limit of the Higgs/higgsino/gauge/gaugino sectors holds
if and only if  Eqs.~(\ref{f_ESL}), (\ref{calY_ESL}) and (\ref{calZ_ESL}) hold.
If Eqs.~(\ref{f_ESL}) and (\ref{calZ_ESL}) hold
but Eq.~(\ref{calY_ESL}) does not,
then supersymmetry is softly broken
due to the quadratic terms of the Higgs potential.
The above results fully resolves the question of the basis-invariant
form for the supersymmetric limit of the THDM.

\subsection{\label{sec:preferred}A preferred basis in the MSSM}

The gaugino-higgsino-Higgs interactions provide a means for defining a \textit{quasi-physical}
choice of basis.  In this context, a quasi-physical basis is one in which
the Higgs fields are invariant, up to a possible rephasing of one of the Higgs fields, under
U(2) transformations.  That is, the coefficients of the Higgs potential
are either U(2)-invariants or pseudo-invariants.

There are four possible independent quasi-physical bases, corresponding to the
four normalized gaugino-higgsino-Higgs couplings, $\hat f_U^a$, $\hat f_D^a$, $\hat f_U^{\prime\,a}$
and $\hat f_D^{\prime\,a}$.   Each basis is defined by imposing the condition that
one of the two components of the corresponding coupling vanishes, while setting
the non-vanishing component to unity.
This defines the quasi-physical basis up to an arbitrary rephasing
of the Higgs field that lies in the direction of the vanishing
component of $f$.  The latter
is a pseudo-invariant field, whereas the Higgs field that lies in the direction of
the non-vanishing component of $f$ is a U(2)-invariant field.\footnote{In the most general 2HDM after
spontaneous symmetry breaking, the so-called Higgs basis in which $\hat{v}^a=(1\,,\,0)$ is an example
of a quasi-physical basis.  The Higgs fields in this basis, $H_1\equiv \hat{v}_a^*\Phi_a$ and
$H_2\equiv \epsilon_{ba} \hat{v}_b \Phi_a$, are respectively invariant and pseudo-invariant with
respect to U(2) basis transformations.  See Refs.~\cite{LS,BS,BLS,DavHab,Haber,oneil} for details.}

The four quasi-physical bases and their corresponding Higgs fields are:
\beqa
\mathcal{B}_U: &\quad& \textrm{defined by}~\hat
f_U^a=(0\,,\,1)\,,\qquad \textrm{Higgs fields}:~~(\mathcal{H}_U\,,\,
H_U)\,,\label{bu}\\
\mathcal{B}_D: &\quad& \textrm{defined by}~\hat
f_D^a=(1\,,\,0)\,,\qquad \textrm{Higgs fields}:~~(H_D\,,\,
\mathcal{H}_D)\,,\label{bd}\\
\mathcal{B}^\prime_U: &\quad &\textrm{defined by}~\hat
f^{\prime\,a}_U=(0\,,\,1)\,,\qquad
\textrm{Higgs fields}:~~(\mathcal{H}^\prime_U\,,\,H^\prime_U)\,,\label{bup}\\
\mathcal{B}^\prime_D: &\quad &\textrm{defined by}~\hat
f^{\prime\,a}_D=(1\,,\,0)\,,\qquad \textrm{Higgs
  fields}:~~(H^\prime_D\,,\,\mathcal{H}^\prime_D)\,,\label{bdp}
\eeqa
where the fields denoted by (calligraphic) Roman fonts are (pseudo-)invariant with respect
to U(2) basis transformations.  The Higgs fields with a $U$ ($D$) subscript are hypercharge
$+1$ ($-1$) fields.  The coefficients of the scalar potential
in the quasi-physical basis are
easily constructed.  For example, in basis $\mathcal{B}_D$,
\beqa
Y_{D1}&\equiv& \hat f_D^{a\ast} \hat f_D^b Y_{ab}\,,\qquad \qquad\quad\,\,
Y_{D2}\equiv \hat g_D^{a\ast} \hat g_D^b Y_{ab}\,,\qquad  \label{cov1}\\
Y_{D3}&\equiv &\hat f_D^{a\ast} \hat g_D^b Y_{ab}\,,\qquad\qquad\quad\,\,
Z_{D1}\equiv \hat f_D^{a\ast} \hat f_D^b \hat f_D^{c\ast} \hat f_D^d Z_{abcd}\,,\label{cov2}\\
Z_{D2}&\equiv& \hat g_D^{a\ast} \hat g_D^b \hat g_D^{c\ast} \hat g_D^d Z_{abcd}\,,\qquad
Z_{D3}\equiv \hat f_D^{a\ast} \hat f_D^b \hat g_D^{c\ast} \hat g_D^d Z_{abcd}\,,\label{cov3}\\
Z_{D4}&\equiv& \hat g_D^{a\ast} \hat f_D^b \hat f_D^{c\ast} \hat g_D^d Z_{abcd}\,,\qquad
Z_{D5}\equiv \hat f_D^{a\ast} \hat g_D^b \hat f_D^{c\ast} \hat g_D^d Z_{abcd}\,,\label{cov4}\\
Z_{D6}&\equiv& \hat f_D^{a\ast} \hat f_D^b \hat f_D^{c\ast} \hat g_D^d Z_{abcd}\,,\qquad
Z_{D7}\equiv \hat f_D^{a\ast} \hat g_D^b \hat g_D^{c\ast} \hat g_D^d Z_{abcd}\,.\label{cov5}
\eeqa
That is, starting in a generic basis with a Higgs potential given by \eq{VH1},
one can always transform to the basis $\mathcal{B}_D$ with a Higgs potential
given by:
\beqa
V_H
&=&
Y_{D_1}H_D^\dagger H_D + Y_{D_2} \mathcal{H}_D^\dagger \mathcal{H}_D
+\left[ Y_{D_3} H_D^\dagger \mathcal{H}_D + \textrm{h.c.} \right]
\nonumber\\[6pt]
&&
+ \tfrac{1}{2} Z_{D_1} (H_D^\dagger H_D)^2
+ \tfrac{1}{2} Z_{D_2} (\mathcal{H}_D^\dagger \mathcal{H}_D)^2
+ Z_{D_3} (H_D^\dagger H_D) (\mathcal{H}_D^\dagger \mathcal{H}_D)
+ Z_{D_4}(H_D^\dagger \mathcal{H}_D) (\mathcal{H}_D^\dagger H_D)
\nonumber\\[6pt]
&&
+ \left[
\tfrac{1}{2} Z_{D_5} (H_D^\dagger \mathcal{H}_D)^2
+ Z_{D_6} (H_D^\dagger H_D) (H_D^\dagger \mathcal{H}_D)
+ Z_{D_7}(\mathcal{H}_D^\dagger \mathcal{H}_D) (H_D^\dagger \mathcal{H}_D)
+ \textrm{h.c.}
\right].
\label{VHD1}
\eeqa
The parameters $Y_{D1}, Y_{D2}, Z_{D1}, Z_{D2}, Z_{D3}$ and $Z_{D4}$
are manifestly real U(2)-invariants, whereas the parameters
$Y_{D3}, Z_{D5}, Z_{D6}$ and $Z_{D7}$ are (potentially) complex
pseudo-invariants with respect to U(2) basis transformations.
The physical Higgs couplings and masses of the theory can be expressed
in terms of the invariant parameters and invariant combinations of
the pseudo-invariant parameters.

The discussion above is completely general.  But, suppose we impose
supersymmetry on the Higgs-higgsino-gaugino system.  Supersymmetry
imposes a very strong constraint on the quasi-physical bases,
\beq \label{susyb}
\mathcal{B}_U=\mathcal{B}_D=\mathcal{B}^\prime_U=\mathcal{B}^\prime_{D}\,.
\eeq
That is, supersymmetry picks out
a physically meaningful basis, namely the SUSY basis, in which the gaugino-higgsino-Higgs couplings take
the form given by \eq{spbasis}.  \Eq{susyb} impose
covariant relations among the normalized gaugino-higgsino-Higgs couplings:
\beq \label{susyfg}
\hat g_D^a=\hat g_D^{\prime\,a}=e^{i\eta}\hat f_U^a=e^{i\eta}\hat f_U^{\prime\,a}\,,\qquad\qquad
\hat g_U^a=\hat g_U^{\prime\,a}=-e^{i\eta}\hat f_D^a=-e^{i\eta}\hat f_D^{\prime\,a}\,,
\eeq
where $e^{i\eta}$ is a pseudo-invariant quantity [i.e., $e^{i\eta}\to (\det~U)^{-1}e^{i\eta}$ under
a U(2) transformation] that is
equal to $1$ in the SUSY basis.  \Eq{susyfg} implies that the invariant and
pseudo-invariant Higgs fields defined in \eqst{invH1}{invH5} are related in the supersymmetric limit,
\beqa
H_U=H_U^\prime&=&e^{i\eta}\widetilde{\mathcal{H}}_D=e^{i\eta}\widetilde{\mathcal{H}}_D^\prime\,,\label{mssmhu}\\
H_D=H_D^\prime&=&e^{-i\eta}\widetilde{\mathcal{H}}_U=e^{-i\eta}\widetilde{\mathcal{H}}_U^\prime\,,\label{mssmhd}
\eeqa
for the hypercharge $+1$ and hypercharge $-1$ MSSM Higgs fields, respectively.
If we now insert \eq{susyfg} into
\eqst{cov1}{cov5}, we find that the coefficients of the Higgs potential
in the SUSY basis are $\mathcal{Y}_{DD},\mathcal{Y}_{UU},e^{i\eta}\mathcal{Y}_{UD},
\mathcal{Z}_{UU,UU},\mathcal{Z}_{DD,DD},\mathcal{Z}_{DD,UU},\mathcal{Z}_{DU,UD},e^{2i\eta}\mathcal{Z}_{DU,DU},
e^{i\eta}\mathcal{Z}_{DD,DU},e^{i\eta}\mathcal{Z}_{UU,UD}$, subject to the conditions
of \eqs{calY_ESL}{calZ_ESL}.
Note that when the conditions of \eqs{calY_ESL}{calZ_ESL} are imposed,
the phase $\eta$ drops out completely.
Indeed, the nonzero Higgs potential parameters in the SUSY basis,
$\mathcal{Y}_{DD},\mathcal{Y}_{UU},\mathcal{Z}_{UU,UU},\mathcal{Z}_{DD,DD},\mathcal{Z}_{DD,UU},\mathcal{Z}_{DU,UD}$
are physical observables that are real and invariant with respect to U(2)-basis transformations

The existence of a physical SUSY basis is not surprising.  Consider the
higgsino Majorana mass term, which arises from a supersymmetric term in
the MSSM superpotential,
%
\begin{equation}
\mu \epsilon_{ij} \psi^i_{H_D} \psi^j_{H_U}.
\label{fix}
\end{equation}
A THDM basis transformation mixes $\Phi_1$ and $\Phi_2$, which corresponds
to a transformation between the MSSM Higgs fields,
$H_U$ with $H_D^\dagger$.
But the Higgs fields belong to Higgs superfields whose scalar and
fermionic components are $(H_U, \psi_{H_U})$ and $(H_D, \psi_{H_D})$.
As a result, one cannot apply a general U(2)-basis transformation
to the full Higgs supermultiplets consistent with supersymmetry.
Thus,  Eq.~(\ref{fix}) effectively defines a preferred basis for
$\Phi_1$ and $\Phi_2$ within the MSSM.
This is why the SUSY basis has a physical significance.

\subsection{\label{sec:tanb}Invariant $\boldsymbol{\tan\beta}$-like parameters}

Until this subsection, we have made no assumptions about the nature of the Higgs vacuum.  In
the supersymmetric limit of the MSSM, the minimum of the tree-level potential
resides at zero field and there is no electroweak symmetry breaking.
After dimension-two soft-supersymmetry-breaking terms are included,
it becomes possible
to break the SU(2)$\times$U(1) symmetry of the scalar potential down to U(1)$_{\rm EM}$.
In the generic basis, we define vacuum expectation values,
\beq
\langle \Phi^0_a\rangle  = v_a\equiv v\hat v_a\,,
\eeq
where $\hat v_a$ is a complex unit vector in the two-dimensional Higgs flavor space and $v\simeq 246~{\rm GeV}$.
In the MSSM, the parameter $\tan\beta$ is defined in the SUSY basis,\footnote{In the SUSY basis,
the overall phase of the Higgs fields can be chosen such that $v_U$ and $v_D$ are real and positive.}
\beq
\tan\beta\equiv \frac{v_U}{v_D}\,.
\eeq
However, in the most general THDM, $\tan\beta$ is a basis-dependent concept
that does not correspond to any physical observable.  In Ref.~\cite{oneil}, a number
of invariant parameters connected to the Higgs-fermion Yukawa couplings
are introduced that play the role of $\tan\beta$ in constrained THDMs.
Here, we shall indicate how to define $\tan\beta$-like parameters
associated with the gaugino-higgsino-Higgs interactions.

The basic idea is to define $\tan\beta$ as the ratio of vacuum expectation values
in the quasi-physical bases introduced in \eqst{bu}{bdp}.  One can then define four invariant
$\tan\beta$-like parameters,\footnote{The signs in \eqs{tb1}{tb2} have
been conveniently chosen so that no extraneous signs appear in
the SUSY limit [cf.~\eq{susytb}].}
\beqa
\tan\beta_U&\equiv& -e^{-i\eta}\frac{\langle H_U^0\rangle}{\langle \mathcal{H}_U^0\rangle}
=-e^{-i\eta}\frac{\hat f_U^{a\ast}\hat v_a}{\hat g_U^{a\ast}\hat v_a}\,,
\qquad\qquad
\tan\beta_D\equiv \,e^{-i\eta}\frac{\langle \mathcal{H}_D^0\rangle}{\langle H_D^0\rangle}
=e^{-i\eta}\frac{\hat g_D^{a}\hat v_a^\ast}{\hat f_D^{a}\hat v_a^\ast}\,,\label{tb1}\\
\tan\beta^\prime_U&\equiv& -e^{-i\eta}\frac{\langle H_U^{\prime\,0}\rangle}
{\langle\mathcal{H}_U^{\prime\,0}\rangle}= -e^{-i\eta}\frac{\hat f_U^{\prime\,a\ast}\hat v_a}
{\hat g_U^{\prime\,a\ast}\hat v_a}\,,
\qquad\qquad\!\!\!
\tan\beta^\prime_D\equiv \,e^{-i\eta}\frac{\langle\mathcal{H}_D^{\prime\,0}\rangle}
{\langle H_D^{\prime\,0}\rangle}=e^{-i\eta}\frac{\hat g_D^{\prime\,a}\hat v_a^\ast}
{\hat f_D^{\prime\,a}\hat v_a^\ast}\,.\label{tb2}
\eeqa
Without the factors of $e^{i\eta}$, the corresponding $\tan\beta$-like parameters
are complex pseudo-invariants, whose magnitudes are basis-independent.
The interpretation of these
$\tan\beta$-like parameters is simplest in the Higgs basis, which is defined by $\hat v=(1\,,\,0)$.
In the Higgs basis, the parameters in \eqs{tb1}{tb2} are
(up to an overall phase) simply the ratios of gaugino-higgsino-Higgs couplings.
One can now investigate the limit
in which all dimension-four couplings respect supersymmetry.   In this case,
the invariant and pseudo-invariant Higgs fields are constrained by \eqs{mssmhu}{mssmhd},
in which case,
\beq \label{susytb}
\tan\beta\equiv\frac{\langle H_U^0\rangle}{\langle H_D^0\rangle^\ast}=\tan\beta_U=(\tan\beta_D)^\ast
=\tan\beta_U^\prime=(\tan\beta_D^\prime)^*\,,\qquad \text{[SUSY limit]}\,.
\eeq
Corrections to the above relations at the few percent level are expected at the one-loop level due to
supersymmetry-breaking effects that enter into the loops.
In principle, one could extract the $\tan\beta$ parameters of
\eqs{tb1}{tb2} from
collider data (assuming gaugino-higgsino-Higgs couplings could be measured with sufficient accuracy),
and check to see whether \eq{susytb} holds.






\section{\label{sec:conclusions}Conclusions}
The THDM includes the Higgs sector of the MSSM
as a particular case.
In the THDM only quantities that are invariant under
a unitary basis change, $\Phi_a\to U_{ab}\Phi_b$ (where $a,b=1,2$ and
$U$ is an arbitrary unitary matrix) can have physical meaning.
Thus,
it is natural to look for basis invariant definitions of
the supersymmetric limit.
We have shown that there is no basis-invariant definition of
a supersymmetric THDM based on invariants defined exclusively
in terms of parameters of the scalar Higgs potential.
We have constructed basis invariant
probes of the supersymmetric limit and soft-supersymmetry-breaking
by employing both the Higgs potential and the gaugino-higgsino-Higgs interactions.
We also observe that the usual basis choice of
$H_U$ and $H_D$ in the MSSM does have a physical significance,
due to the $\mu$-term of the MSSM superpotential [cf.~Eq.~(\ref{fix})].

Finally, we addressed the physical significance of the parameter
$\tan\beta$.  In the MSSM, the tree-level parameter $\tan\beta$
is well-defined because it is defined in terms of vacuum expectation
values of the neutral Higgs fields, which are given in a physical
basis.  However, in the most general THDM Higgs sector
coupled to new particles with electroweak quantum numbers
that coincide with the gauginos and higgsinos of the MSSM, the
parameter $\tan\beta$, which is defined in terms of a ratio of Higgs vacuum
expectation values, is a basis dependent quantity and hence unphysical.
We have shown that four invariant $\tan\beta$-like parameters can be
defined that are basis-independent and hence physical.  One can verify
that in the supersymmetric limit, these four invariant parameters
coincide (at tree-level) and are equal to the $\tan\beta$ parameter of
the MSSM.


\begin{acknowledgments}
The work of P.M.F. is supported in part by the Portuguese
\textit{Funda\c{c}\~{a}o para a Ci\^{e}ncia e a Tecnologia} (FCT)
under contract PTDC/FIS/70156/2006. The work of H.E.H. is
supported in part by the U.S. Department of Energy, under grant
number DE-FG02-04ER41268. The work of J.P.S. is supported in
part by FCT under contract CFTP-Plurianual (U777) and
under project CERN/FP/109305/2009,
and by the EU RTN project Marie Curie: MRTN-CT-2006-035505.
H.E.H. is most grateful for the kind hospitality and support of the
Centro de F\'{\i}sica Te\'orica e Computacional at Universidade de
Lisboa during his visit to Lisbon.  He also acknowledges
conversations with Sacha Davidson and John Mason, which inspired
a number of ideas presented in this paper.

\end{acknowledgments}

\end{document}